%% file: main.tex
\pdfoutput=1
\documentclass[
aps,%
prb,%
twocolumn,%
superscriptaddress,%
citeautoscript,%
amsmath,%
amssymb,%
amsfonts,%
showkeys,%
reprint,%
floatfix%
]{revtex4-2}

\usepackage[utf8]{inputenc}
\usepackage{graphicx}
\graphicspath{ {./figures/} }
\usepackage{booktabs}
\usepackage{dcolumn}
\usepackage{bm}
\usepackage{amssymb}
\usepackage{amsmath}
\usepackage{color}
\usepackage{xcolor}
\usepackage{xparse}
\usepackage{enumitem}
\usepackage{xkvltxp}
\usepackage{xcolor}
\usepackage{comment}
\usepackage{ifthen}

\usepackage[colorlinks,linkcolor=blue,citecolor=blue,urlcolor=blue]{hyperref}
\usepackage[all]{hypcap}

\newcommand{\Q}{\mathbf{Q}}

\newcommand{\ket}[1]{
  \mathchoice%
  {\left|#1\right\rangle}       
  {|#1\rangle}                  
  {|#1\rangle}                  
  {|#1\rangle}                  
}
\newcommand{\braket}[2]{\left\langle#1\middle|#2\right\rangle}
\newcommand{\matrixel}[3]{\left\langle#1\middle|#2\middle|#3\right\rangle}

\def\tA2{^{3}\!A_2}
\def\sA1{^{1}\!A_1}

\newcommand*{\pgroup}[2]{\ensuremath{{#1}_{#2}}}

\NewDocumentCommand{\gwf}{s O{i} O{n}}{
  \IfBooleanTF#1
  {\Psi_{#3}^{#2}}
  {\ket{\Psi_{#3}^{#2}}}
}
\NewDocumentCommand{\ggwf}{s O{i} O{n}}{
  \IfBooleanTF#1
  {\Phi_{#3}^{#2}}
  {\ket{\Phi_{#3}^{#2}}}
}
\NewDocumentCommand{\wf}{O{i}}{\ket{\Phi_{#1}}}
\NewDocumentCommand{\vwf}{s O{i} O{n} O{} O{}}{
  \chi_{#2;#3}^{#4}
  \IfBooleanTF#1
  {}
  {(\Q_{#5})}
}

\newboolean{draft}
\setboolean{draft}{true}
\NewDocumentCommand\todo{m}%
{%
  {\color{blue} (TODO: #1)}%
}
\NewDocumentCommand\change{om}%
{%
  \ifthenelse{\boolean{draft}}{
    \IfNoValueTF{#1}{}%
    {%
      {\color{gray}[#1]}
    }%
    {\color{orange}#2}%
  }%
  {#2}%
}

\begin{document}


\title{Optical lineshapes for orbital singlet to doublet transitions
  in a dynamical Jahn--Teller system: the NiV$^{-}$ center in diamond}

\author{Rokas Silkinis}
\author{Vytautas \v{Z}alandauskas}
\affiliation{Center for Physical Sciences and Technology (FTMC),
  Vilnius LT-10257, Lithuania}

\author{Gerg\H{o} Thiering}
\affiliation{Institute for Solid State Physics and Optics,
  HUN-REN Wigner Research Centre for Physics,
  P.O.~Box~49, H-1525 Budapest, Hungary}

\author{Adam Gali}
\affiliation{Institute for Solid State Physics and Optics,
  HUN-REN Wigner Research Centre for Physics,
  P.O.~Box~49, H-1525 Budapest, Hungary}
\affiliation{Department of Atomic Physics, Institute of Physics,
  Budapest University of Technology and Economics,
  M\H{u}egyetem rkp.~3,
  H-1111 Budapest, Hungary}
\affiliation{MTA-WFK Lend{\"u}let ``Momentum''
  Semiconductor Nanostructures Research Group,
  PO.~Box~49, H-1525 Budapest, Hungary}

\author{Chris G. Van de Walle}
\affiliation{Materials Department, University of California,
  Santa Barbara, CA~93106-5050, USA}

\author{Audrius Alkauskas}
\thanks{Deceased}
\author{Lukas Razinkovas}
\email{lukas.razinkovas@ftmc.lt}
\affiliation{Center for Physical Sciences and Technology (FTMC),
  Vilnius LT-10257, Lithuania}



\begin{abstract}
  We apply density functional theory to investigate interactions between
  electronic and vibrational states in crystal defects with multi-mode dynamical
  Jahn--Teller (JT) systems. Our focus is on transitions between orbital singlet
  and degenerate orbital doublet characterized by
  $E \otimes (e \oplus e \oplus \cdots)$ JT coupling, which frequently occurs in
  crystal defects that are investigated for applications in quantum information
  science. We utilize a recently developed methodology to model the
  photoluminescence (PL) spectrum of the negatively charged split nickel-vacancy
  center (NiV$^-$) in diamond, where JT-active modes significantly influence
  electron--phonon interactions. Our results validate the effectiveness of the
  methodology in accurately reproducing the observed 1.4~eV PL lineshape. The
  strong agreement between our theoretical predictions and experimental
  observations reinforces the identification of the 1.4~eV PL center with the
  NiV$^-$ complex. This study highlights the critical role of JT-active modes in
  affecting optical lineshapes and demonstrates the power of advanced techniques
  for modeling optical properties in complex systems with multiple JT-active
  frequencies.
\end{abstract}


\maketitle


\section{Introduction\label{sec:intro}}

Electronic degeneracy occurs in molecular structures or point defects
in solids with highly symmetric atomic configurations, leading to the
intricate coupling between electronic and vibrational degrees of
freedom known as the Jahn--Teller (JT)
effect~\cite{jahn1937,opik1957studies,longuet1958,bersuker2012}. If
the coupling is not too strong, it has a dynamical aspect, which leads
to rapid dynamic distortions along symmetry-breaking directions. This
phenomenon is called the dynamical Jahn--Teller (DJT)
effect~\cite{longuet1958}.

The dynamic interplay between electronic and ionic degrees of freedom
can influence specific, measurable properties of the electronic
system~\cite{ham1965}. Furthermore, the closely intertwined electronic
and vibrational components can generate a distinct array of vibronic
states, giving rise to specific features within emission or absorption
spectra. The emergence of such spectral attributes was first discussed
in the pioneering paper on the DJT effect by Longuet-Higgins
\textit{et~al.}~\cite{longuet1958}. This research examined the
lineshape of the orbital singlet $A$ to orbital doublet $E$ transition
in an $E \otimes e$ DJT system. They used an effective
single-degenerate-mode model to depict the motion along a
symmetry-breaking direction, which dynamically couples electronic
states. In this fundamental description of the JT effect, they
demonstrated that the lineshapes for $A \to E$ transitions exhibit two
intensity peaks, a hallmark of a JT-active system. While this
single-mode model correctly describes the general features of simple
DJT systems, it raises the question of its precision in characterizing
systems with many JT-active modes. In her theoretical work,
M.C.M.~O'Brien~\cite{obrien1972} showed that with strong vibronic
coupling, the issue of many frequencies can be approximated by an
effective model of a single-degenerate mode. She later affirmed that
this simplification efficiently describes the optical lineshapes of
the two-degenerate-mode system
$E \otimes (e \oplus e)$~\cite{obrien1980}. However, the effectiveness
of the single-effective mode method is not guaranteed in cases of weak
coupling; a multi-mode approach is necessary, as demonstrated by
studies simulating the photoemission spectrum of
$\mathrm{C}_{60}^{-}$~\cite{Gunnarsson1995,Iwahara2010}.

Recent advancements in exploring crystal defects as quantum systems
for technological applications have sparked a renewed interest in the
DJT effect within color
centers~\cite{abtew2011dynamic,thiering2017ab,Zhang2018,thiering2018,thiering2019eg,thiering2021niv}.
A primary challenge in analyzing the DJT effect in such systems is the
electron--phonon coupling to a continuum of vibrational frequencies.
Therefore, theoretical investigations often employ the
single-effective mode model to explore spectral and magnetic
properties. In addition, when modeling optical sidebands in DJT
systems involving many vibrational modes, the standard Huang--Rhys
(HR) theory, developed for adiabatic states, is typically employed to
capture contributions from JT-active
modes~\cite{Zhang2018,thiering2018,thiering2021niv}. A recent
study~\cite{razinkovas2021} highlighted the limitations of both the
single-mode model and HR theory in accurately simulating optical
lineshapes by examining the absorption lineshape of the $A \to E$
triplet transition in the negatively charged nitrogen-vacancy (NV)
center in diamond. In that work, some of the present authors developed
a rigorous methodology based on density functional theory (DFT)
calculations to overcome these limitations and accurately model
multi-mode $E\otimes (e \oplus e \oplus \cdots)$ DJT
systems~\cite{razinkovas2021}.

In this paper, we apply this recently developed methodology to elucidate the
photoluminescence (PL) spectrum of the negatively charged split nickel-vacancy
center (NiV$^-$) in diamond~\cite{thiering2021niv}. In the previous application
of the method to the absorption spectrum of the negatively charged NV center in
diamond~\cite{razinkovas2021}, the JT coupling was minor relative to the
contribution from electron--phonon coupling involving symmetry-preserving $a_1$
modes. While the theory did allow for obtaining the correct weight distribution
in the absorption lineshape, the DJT effect did not yield any sharply visible
spectral features from the JT-active modes. In the NiV$^-$ case, however, the
influence of JT-active modes with $e_g$ symmetry is more than three times
greater than that of symmetry-preserving $a_{1g}$ modes, leading to noticeable
JT-specific features in the optical lineshape that we are able to reproduce
using our methodology. By comparing optical lineshapes derived from standard HR
theory applied to JT-active modes against those obtained through rigorous
treatment of DJT multi-mode systems, we demonstrate that HR theory alone cannot
adequately explain the observed spectral features.

\section{NiV$^-$ center in diamond}

\begin{figure}
  \centering
  \includegraphics[width=0.95\linewidth]{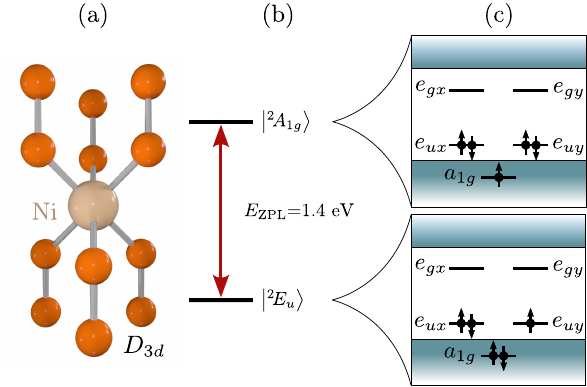}
  \caption{(a) Atomic structure of the split nickel-vacancy (NiV$^-$) center in
    diamond. (b) Electronic level diagram of NiV$^{-}$. (c) Electronic
    configurations in the single-electron picture of $m_s=1/2$ spin sublevels of
    the ground $\ket{^2\!E_{uy}}$ and excited $\ket{^2\!A_{1g}}$
    single-determinant wave functions.\label{fig:niv}}
\end{figure}

Nickel is a common impurity in diamond produced under high-pressure and
high-temperature conditions, creating identifiable features that can be detected
using optical and electron paramagnetic resonance
methods~\mbox{\cite{collins1983,yelisseyev2007,zaitsev2013,nadolinny2017}}. A
notable fingerprint of the nickel impurity is the 1.4~eV PL line~\cite{collins1983},
which is associated with the NiV$^-$ center~\cite{thiering2021niv}.
The atomic configuration of this system is
depicted in Fig.~\ref{fig:niv}(a). The impurity ion resides precisely at the
inversion center between two adjacent vacancies, resulting in the point-group
symmetry of the defect being \pgroup{D}{3d}, displaying spin-doublet
configurations in both the ground and optically excited
states~\cite{thiering2021niv} [Fig.~\ref{fig:niv}(b)].

The molecular orbital model for this center can be constructed by considering
the presence of six carbon dangling bonds surrounding the nickel atom and the
contribution of five $3d$ atomic orbitals of Ni~\cite{thiering2021niv}. This
atomic basis results in a total of 11 molecular orbitals. In the ground state,
seven orbitals constitute filled closed shells located deep within the valence
band. The highest energy orbital among these is $a_{1g}$
[Fig.~\ref{fig:niv}(c)], which is essential in the considered optical
transition. The remaining four orbitals form two degenerate doublets with $e_u$
and $e_g$ symmetry within the band gap. The $e_u$ orbitals are filled with three
electrons, forming an orbital doublet with $E_u$ symmetry. This state can be
described by a single determinant wave function, as shown in
Fig.~\ref{fig:niv}(c). This $E_u$ electronic state constitutes an
$E_u\otimes (e_g\oplus e_g\oplus \cdots )$ DJT system where $e_g$ symmetry
vibrational modes couple the components of degenerate electronic states
($E_{ux}$ and $E_{uy}$).\looseness=-1

In contrast, the optically excited state is an orbital singlet with
$A_{1g}$ symmetry. In this state, one of the spin minority $a_{1g}$
electrons is promoted to the $e_{u}$ orbital, resulting in a
single-determinant wave function depicted in Fig.~\ref{fig:niv}(c).

When considering the electron--phonon interaction during an optical
${^{2}\!A_{1g}} \to {^{2}\!E_{u}}$ transition, the symmetry-breaking JT-active vibrational
modes of $e_g$ symmetry and the symmetry-preserving modes of $a_{1g}$ symmetry
participate in the transition. The participation of the $a_{1g}$ symmetry modes
can be described using the customary adiabatic HR theory~\cite{alkauskas2014}.

The arrangement of optically active states in NiV$^-$ is formally
equivalent to a generalized problem of
$A$-to-$E\otimes (e\oplus e \oplus \cdots)$
transition~\cite{longuet1958,obrien1980}, often observed in trigonal
and octahedral symmetry systems. Therefore, in the subsequent theoretical
description, the electronic $E_u$ and $A_{1g}$, along with the
vibrational $e_g$ and $a_{1g}$ symmetries, will be denoted by $E$,
$A$, $e$, and $a$, emphasizing the applicability of theory across
systems with similar symmetry.

\section{Vibronic broadening of optical emission}

The theory for vibronic broadening of optical lineshapes for $A \to E$
transitions and the analysis of multi-mode $E \otimes (e \oplus e \oplus \cdots )$
JT systems is detailed in Ref.~\cite{razinkovas2021,razinkovasthesis}. Here, we
briefly review the formulas describing the PL lineshape of the $A$-to-$E$
transition; we refer the reader to the original references for a complete
derivation.

\subsection{Vibronic states}

As the initial state in the emission process (labeled by~$i$)
is an excited state characterized as an orbital singlet, we can
express its vibronic wave function in an adiabatic form:
\begin{equation}
\ket{\Psi_{i;kl}} = \vwf[i][l][a][a] \vwf[i][k][e][e] \ket{A},
\label{WF1}
\end{equation}
where $\vwf[i][l][a][a]$ and $\vwf[i][k][e][e]$ are the harmonic components of
the vibrational wave function, corresponding to the $a$- and $e$-symmetry modes,
with the quantum numbers $l$ and $k$ labeling different harmonic excitations.
$\mathbf{Q}$~describes the set of symmetry-adapted normal coordinates. The term
$\ket{A}$ denotes the electronic wave function, which in the static adiabatic
approximation depends only on the electronic degrees of freedom.

The vibronic wave function within the degenerate electronic manifold of the
final (ground) state, labeled by $f$, corresponding to the JT-active state, is
expressed as
\begin{equation}
\ket{\Psi_{f;nm}} = \vwf[f][n][a][a]\ket{\Phi_{f;m}},
\label{WF2}
\end{equation}
where $\vwf[f][n][a][a]$ represents the harmonic states of the modes with
$a$-symmetry. The JT component of the problem is captured by
\begin{equation}
\ket{\Phi_{f;m}} =
\vwf[f][m][e;x][e]\ket{E_{x}} +
\vwf[f][m][e;y][e]\ket{E_{y}},
\label{eq:vibronic_term}
\end{equation}
where $\ket{E_{x}}$ and $\ket{E_{y}}$ represent a basis that
transforms as Cartesian irreducible representations of the degenerate
doublet of electronic states, and $\vwf[f][m][e;\alpha][e]$ are
corresponding ionic prefactors that must be determined by solving the
JT problem~\cite{razinkovas2021}. The quantum numbers $n$ and $m$
respectively label the harmonic and vibronic excitations of the
system.

\subsection{Emission lineshape}

Within the zero-temperature limit ($T=0$~K) of the Franck--Condon approximation,
the normalized luminescence intensity $L(\hbar\omega)$ is expressed as~\cite{razinkovas2021}
\begin{eqnarray}
  \label{eq:P}
  L(\hbar\omega)=C\omega^{3}A(\hbar\omega),
\end{eqnarray}
where $C$ is a normalization constant, and $A(\hbar\omega)$ is the spectral
function that holds the information about the shape of the phonon sideband.

In the case of $A\to E$ transition, the spectral function is described by a
specific mathematical expression~\cite{razinkovas2021}:
\begin{eqnarray}
  A(\hbar\omega)
  =
  \int
  A_{a}(\hbar\omega-\hbar\omega')
  A_{e}(\hbar\omega')
  \mathrm{d}(\hbar\omega').
\label{conv}
\end{eqnarray}
This equation represents a convolution of two spectral functions
corresponding to $a$- and $e$-symmetry modes:
\begin{align}
  A_{a}(\hbar\omega) =
  &
    \sum_n
    \left|
    \braket{\vwf*[f][n][a][a]}{\vwf*[i][0][a][a]}
    \right|^2
    \delta
    \left(
    E_{\mathrm{ZPL}} + \varepsilon^{a}_{fn} - \varepsilon^{a}_{i0} -\hbar\omega
    \right),
    \notag
  \\
  A_e(\hbar\omega) =
  &
    \sum_m
    \left[
    \bigl|
    \bigl\langle{\vwf*[f][m][e;x][e]}\big|{\vwf*[i][0][e][e]}\bigr\rangle
    \bigr|^2 +
    \bigl|
    \bigl\langle{\vwf*[f][m][e;y][e]}\big|{\vwf*[i][0][e][e]}\bigr\rangle
    \bigr|^2
    \right]
    \notag \\[-0.2em]
  & \quad \times
    \delta(\varepsilon^{e}_{g0} - \varepsilon^e_{gr} - \hbar\omega) \, .
    \label{eq:spectral_f}
\end{align}
In this formulation, the terms $\varepsilon^{\gamma}_{\alpha n}$ represent the
energy eigenvalues of the vibrational/vibronic quantum mechanical level $n$
within the electronic state $\alpha$, corresponding to the symmetry $\gamma$ of
the vibrational degrees of freedom. $E_{\mathrm{ZPL}}$ denotes the zero-phonon
line (ZPL) energy.

\subsection{Coupling to $a$ modes}

To simplify the estimation of overlap integrals
$\langle\vwf*[f][n][a][a]|\vwf*[i][0][a][a]\rangle$ entering
$A_{a}(\hbar\omega)$, we employ the equal-mode
approximation~\cite{markham1959,razinkovas2021}. This approach assumes that the
vibrational modes of the initial state are identical to those of the final
state, thereby reducing the problem of calculating the vibrational structure of
the ground state alone.
Within this approximation, the evaluation of the spectral function
$A_{a}(\hbar\omega)$ can be achieved using the generating function
approach~\mbox{\cite{lax1952,alkauskas2014,razinkovas2021}}:
\begin{align}
  & A_{a}(\hbar \omega) = \frac{1}{2\pi} \int_{-\infty}^{\infty} e^{i\omega t} G(t) e^{-\gamma |t|}
     \, \mathrm{d}t,
     \notag
  \\
  & G(t)
     =
     \exp\left[
     -\frac{iE_{\mathrm{ZPL}}t}{\hbar} \! - \!
     S_{a} \! + \! \int e^{i\omega t}S_{a}(\hbar \omega)\,\mathrm{d}(\hbar\omega)
  \right]\!\!.
  \label{eq:generating-lum}
\end{align}
The function $G(t)$ is the generating function for luminescence, and parameter
$\gamma$ accounts for additional broadening effects not captured by the theory.
The variable $S(\hbar\omega)=\sum_k S_k \delta(\hbar\omega - \hbar\omega_k)$
is the spectral density of the electron--phonon coupling, where
$S_k = \omega_k\Delta Q_k^2/(2\hbar)$ is the partial HR factor~\cite{Huang1950}
of $a$-symmetry mode $k$, and $\Delta Q_k$ signifies the alteration of the
equilibrium geometry upon optical transition along the direction of vibrational
mode $\boldsymbol{\eta}_{k;\alpha}$ (in mass-weighted form) of frequency
$\omega_k$. We calculate $\Delta Q_k$ by employing the force
$\boldsymbol{F}_{\alpha}$ exerted on atom $\alpha$ of mass $M_\alpha$ as induced
by the electronic transition~\cite{razinkovas2021}:
\begin{align}
  \Delta Q_k = \frac{1}{\omega_k^2}
  \sum_{\alpha}\frac{\mathbf{F}_{\alpha}}{\sqrt{M_{\alpha}}}\boldsymbol{\eta}_{k;\alpha}.
  \label{eq:deltaQk}
\end{align}

\subsection{Coupling to $e$ modes}

In the case of $e$ symmetry modes, the spectral function $A_e(\hbar\omega)$ must
be calculated by explicitly solving the vibronic problem
$\hat{\mathcal{H}} = \hat{\mathcal{H}}_0 + \hat{\mathcal{H}}_{\mathrm{JT}}$ for
states described by Eq.~\eqref{eq:vibronic_term}, where~\cite{bersuker2012}
\begin{align}
  \hat{\mathcal{H}}_{\mathrm{0}} = \mathcal{C}_{z} \sum_{k;\gamma\in \{x, y\}}
  \left(
    -\frac{\hbar^2}{2}\frac{\partial^2}{\partial Q^2_{k\gamma}}
    + \frac{1}{2}  \omega_k^2 Q_{k\gamma}^2
  \right)
\label{eq:Hph}
\end{align}
describes the motion within the harmonic potential and
\begin{equation}
  \hat{\mathcal{H}}_{\mathrm{JT}} = \sum_{k;\gamma\in \{x, y\}}\mathcal{C}_{\gamma} \sqrt{2\hbar\omega_{k}^{3}}K_k Q_{k\gamma}
\label{eq:JT}
\end{equation}
characterizes the linear JT interaction. Here, $\omega_k$ represents
the angular frequencies of vibrations, derived as eigensolutions of the zero-order
Hamiltonian [Eq.~\eqref{eq:Hph}], and $K_k$ denotes the dimensionless vibronic
coupling constants~\cite{obrien1980}. The index $k=1, \dots, N$ encompasses all pairs of degenerate
$e$-symmetry vibrations. The matrices $\mathcal{C}_{\gamma}$ act on orbital
states, and in the Cartesian representation of degenerate orbitals have the
following form~\cite{Ham1968}:
\[
  \mathcal{C}_{x} =
  \begin{pmatrix}
    0 & 1 \\ 1 & 0
  \end{pmatrix},
  \quad
  \mathcal{C}_{y} =
  \begin{pmatrix}
    1 & 0 \\ 0 & -1
  \end{pmatrix},
  \quad
  \mathcal{C}_{z} =
  \begin{pmatrix}
    1 & 0 \\ 0 & 1
  \end{pmatrix}.
\]

Following symmetry arguments from
Refs.~\cite{longuet1958,obrien1980,razinkovas2021}, we represent the
JT Hamiltonian $\hat{\mathcal{H}}_{\mathrm{JT}}$ in the basis of
states of $\hat{\mathcal{H}}_{0}$, which are also eigenstates of the
quasi-angular momentum
$\hat{J} = \hat{J}_{\mathrm{el}} + \hat{J}_{\mathrm{ph}}$. Here,
$\hat{J}_{\mathrm{el}} = \frac{\hbar}{2} \hat{\sigma}_y$ acts on the orbital
component of the wave function, whereas in the Cartesian representation
$\hat{\sigma}_y$ is the Pauli matrix.
$\hat{J}_{\mathrm{ph}} = \mathcal{C}_z \sum_k \mathcal{L}_{k}$ quantifies the
total angular momentum of $e$ symmetry harmonic modes, where
$\mathcal{L}_{k} = \hbar (\hat{n}_{k+} - \hat{n}_{k-})$ represents the angular
momentum operator for a $k$ vibrational doublet. Here,
$\hat{n}_{k\pm} = a^{\dagger}_{k\pm}a_{k\pm}$ defines the number operator for
right- and left-hand phonons, expressed through second quantization operators
$a_{k\pm} = \frac{1}{\sqrt{2}}(a_{kx} \mp ia_{ky})$, with $a_{kx}$ and $a_{ky}$
directly linked to normal modes $Q_{kx}$ and $Q_{ky}$. In this basis, the
eigenstates of $\hat{\mathcal{H}}_{0}$ are denoted by
$|n_{1}l_{1} \dots n_{N}l_{N}; E_{\pm} \rangle$, with orbital wavefunctions
$\ket{E_{\pm}} = \frac{1}{\sqrt{2}}(\ket{E_x} \pm i\ket{E_y})$ having quantum
numbers $j_{\mathrm{el}}=\pm \frac{1}{2}$. Here, $n_{k}=n_{k+}+n_{k-}$
represents the total number of phonons, and $l_{k}$ is the angular momentum
quantum number for each $k$ vibrational doublet. The matrix elements of
$\hat{\mathcal{H}}_0$ and $\hat{\mathcal{H}}_{\mathrm{JT}}$ are then formulated
as follows~\cite{razinkovas2021}:\looseness=-1
\begin{align}
  & \matrixel{n_{1}l_{1},\ldots,n_{N}l_{N};E_{\pm}}
     {{\hat{\mathcal{H}}}_{\mathrm{0}}}
     {n_{1}l{}_{1},\ldots,n_{N}l_{N};E_{\pm}}
     \notag
  \\
  & \quad  = \sum_k \hbar \omega_k \left ( n_k + 1\right),
     \label{eq:matrix_element2}
  \\
  & \matrixel{n'_{1}l'_{1},\ldots,n'_{N}l'_{N};E_{-}}
     {{\hat{\mathcal{H}}}_{\mathrm{JT}}}
     {n_{1}l{}_{1},\ldots,n_{N}l_{N};E_{+}} \notag\\
  & \quad  = \sqrt{2}\sum_{k}K_k\hbar\omega_k\delta_{l'_{k}l_{k}+1}
     \left[\prod_{j\neq k}\delta_{n'_{j}n_{j}}\delta_{l'_{j}l_{j}}\right]
     \notag\\
  & \qquad \times \left[\!\sqrt{\tfrac{n_{k}-l_{k}}{2}}
     \delta_{n'_{k}n_{k}-1}+\sqrt{\tfrac{n_{k}+l_{k}+2}{2}}\delta_{n'_{k}n_{k}+1}\!\right].
     \label{eq:matrix_element}
\end{align}
Since $\hat{J}$ commutes with $\hat{\mathcal{H}}_{\mathrm{JT}}$, the solution
for $\hat{\mathcal{H}}$ can be found separately for each total quantum number
$j = j_{\mathrm{el}} + \sum_{k}l_{k}$, taking the following form:
\begin{align}
  \label{eq:1}
  \ket{\Phi_{f;m}}
  & = \vwf[f][m][e;+][e]\ket{E_{+}} + \vwf[f][m][e;-][e]\ket{E_{-}}
    \notag
  \\
  &
    = \sum_{s=\{{+},{-}\}}\sum_{\mathbf{nl}} C^{s}_{f;m\mathbf{nl}} \ket{n_1l_1\cdots n_Nl_N;E_s},
\end{align}
where $C^{\pm}_{f;m\mathbf{nl}}$ are coefficients obtained through diagonalization.

In this new basis, the bracketed term in Eq.~\eqref{eq:spectral_f} can be
expressed as
$| \langle \vwf*[f][m][e;+][e] | \vwf*[i][0][e][e] \rangle |^2 + | \langle \vwf*[f][m][e;-][e] | \vwf*[i][0][e][e] \rangle |^2$.
To calculate these overlap integrals for $e$-symmetry vibrational modes,
similarly to $a$ modes, we use the equal-mode approximation, which assumes that
the vibrational shapes and frequencies of the $A$ orbital manifold are
well-represented by the zero-order Hamiltonian of the ground state
[Eq.~\eqref{eq:Hph}]. Furthermore, in the zero-temperature limit, these overlaps
are calculated between the zero-phonon state of the excited manifold, denoted as
${\vwf*[i][0][e][e]} = \ket{00\cdots0}$, and all vibronic states of the $E$
manifold. Given that $\sum_{k}l_{k}$ is conserved, only vibronic solutions where
$j=\pm 1/2$ (with $j_{\mathrm{el}}=\pm 1/2$ and $\sum_{k}l_{k}=0$) are relevant
for this analysis.

\section{First-principles methods}

The electronic structure and optical excitation energies of the NiV$^-$ center
in diamond were investigated through spin-polarized DFT. To calculate the
excited state energy and geometry within the framework of Kohn--Sham (KS) DFT, we
employed the delta-self-consistent-field ($\Delta$SCF)
method~\cite{jones1989,hellman2004,kowalczyk2011}. In this approach, an $a_{1g}$
electron from the lower-lying occupied KS orbital was excited to an empty $e_u$
orbital, as illustrated in Fig.~\ref{fig:niv}(c). We utilized the r$^2$SCAN
functional~\cite{r2SCAN}, which combines the numerical efficiency of
rSCAN~\cite{rscan} with the transferable accuracy of SCAN~\cite{scan}. Notably,
this functional has demonstrated excellent performance in capturing the
structural and electronic characteristics of other deep-level
defects in diamond~\cite{scandiamond}. Our calculations were conducted within
$4\times4\times4$ supercells, encompassing 512 atomic sites, with the
Brillouin-zone sampling centered at the $\Gamma$-point. The projector-augmented
wave (PAW) method was employed, utilizing a plane-wave energy cutoff of 600~eV.
These calculations were performed using the Vienna Ab~initio Simulation Package
({\sc VASP})~\cite{vasp}.

\subsection{Zero-order vibrational modes}

Determining vibrational modes within the $E_u$ state, specifically pertaining to
both $a$ and $e$ irreducible representations, requires an approach that excludes
the influence of first-order ($\hat{\mathcal{H}}_{\mathrm{JT}} = 0$) and
higher-order JT couplings. To achieve this separation of contributions, we
employ an \textit{ab initio} methodology, as discussed in detail in
Ref.~\cite{razinkovasthesis}.

In this approach, we consider an electronic configuration characterized by
fractional KS orbital occupation, namely $a_{1g}^2e_{ux}^{1.5}e_{uy}^{1.5}$. This
electronic configuration approximates an ensemble state of two degenerate
configurations ($a_{1g}^2e_{ux}^{1}e_{uy}^{2}$ and
$a_{1g}^2e_{ux}^{2}e_{uy}^{1}$), effectively suppressing all JT interactions
while preserving the inherent geometry associated with \pgroup{D}{3d} symmetry.
Within this configuration, we employ the finite-difference method, as
implemented in the \texttt{Phonopy} software package~\cite{phonopy1,phonopy2}, to compute the
vibrational structure, ensuring that the vibrational modes are well defined with
respect to their irreducible representations.

\subsection{Vibrational structure and relaxation profile in the dilute limit}
\label{sec:vibdil}

Moderately sized supercells, which are computationally amenable to explicit DFT
calculations, pose challenges in accurately capturing the vibrational structure
of defects due to periodic boundary conditions and a limited number of
vibrational degrees of freedom. To overcome these limitations and achieve
high-accuracy and high-resolution lineshapes, we adopt an embedding
methodology~\cite{alkauskas2014,razinkovas2021}. This approach relies on the
short-range character of interatomic interactions and allows for the computation
of vibrational structures within significantly larger supercells.
More details about the embedding methodology are provided in Sec.~1 of the
Supplemental Material~\cite{supp}.

To accurately capture the relaxation profile in the dilute limit, we assess the
relaxation component $\Delta Q_k$ for each vibrational mode. This is achieved by
employing Eq.~\eqref{eq:deltaQk}, which utilizes forces that are already
converged within the explicitly accessible supercell. These forces are then
projected onto the vibrational modes of a system encompassing tens of thousands
of atoms. This methodology captures the participation of low-frequency modes and
provides a detailed description of the vibrational characteristics in these
extensive systems.

\subsection{Vibronic coupling parameters}

In the linear JT theory, the adiabatic potential energy surface of the JT-active
manifold takes the form of a \textit{sombrero} hat~\cite{bersuker2012}. This
surface can be effectively investigated using DFT~\cite{razinkovas2021}.
Allowing the relaxation along an $e$-symmetry direction from the high-symmetry
configuration (obtained using the fractional occupation), one can monitor the
geometry change $\Delta \mathbf{Q}_{\mathrm{JT}}$ that quantifies vibronic
coupling in the case of the linear JT coupling. By projecting
$\Delta \mathbf{Q}_{\mathrm{JT}}$ onto a pair $k$ of $e$-symmetry normal modes,
one can estimate vibronic coupling pertaining to the $k$ vibrational doublet:
\begin{equation}
  K^2_k = \frac{\omega_k \Delta Q_{k}^2}{2\hbar},
\end{equation}
where $\Delta Q_k^2 = \Delta Q_{kx}^2 + \Delta Q_{ky}^2$ describes the
projection of $k$-doublet normal coordinates along the relaxation
$\Delta \mathbf{Q}_{\mathrm{JT}}$ and can be estimated by directly
measuring displacements along each pair of normal coordinates or
through the application of Eq.~\eqref{eq:deltaQk}, which utilizes
force-based calculations.  This method allows the estimation of vibronic
coupling constants using \textit{ab initio} means.

\subsection{Effective modes and diagonalization}

Diagonalizing the Hamiltonian
$\mathcal{H} = \mathcal{H}_0 + \mathcal{H}_{\text{JT}}$, which includes many vibrational modes, presents a significant computational challenge
due to the large size of the matrices. To address this complexity, we employ a
strategy that utilizes a limited set of effective modes~\cite{razinkovas2021}.
Initially, we define the density of the JT coupling as
$K^2(\hbar\omega) = \sum_k K_k^2 \delta(\hbar\omega - \hbar\omega_k)$. We then
approximate this density with
$K^2_{\mathrm{eff}}(\hbar\omega)=\sum_{n=1}^{N_{\mathrm{eff}}} \bar{K}^2_{n}g_{\sigma}(\hbar\bar{\omega}_{n}-\hbar\omega)$,
where $g_{\sigma}$ represents a Gaussian function characterized by a width
$\sigma$. This approximation incorporates $N_{\mathrm{eff}}$ effective
vibrations, each parameterized by a frequency $\bar{\omega}_n$ and a
corresponding vibronic coupling strength $\bar{K}^2_{n}$. The parameters
$\bar{K}^2_n$, $\bar{\omega}_{n}$, and $\sigma$ are optimized to ensure that
$K^2_{\mathrm{eff}}(\hbar\omega)$ closely matches $K^2(\hbar\omega)$. This
approach enables using fewer effective modes
$N_{\mathrm{eff}} \ll N$, making diagonalization more tractable. To ensure the
reliability of the computed spectral functions, we monitor their convergence
as a function of the number of effective modes. Additionally,
when constructing the basis for $\mathcal{H}$, we limit the total number of
excited phonons, $n_{\mathrm{tot}}=\sum_k n_k$, to a predefined threshold and
also track the convergence of our results as this limit increases.

\section{Results and comparison with experiment}

We calculated the optical excitation energy as the energy difference between the
adiabatic potential energy surface minima of the $^{2}\!E_{u}$ and
$^{2}\!A_{1_{g}}$ states. The obtained value of 1.36~eV aligns closely with the
experimental ZPL energy of 1.40~eV and is consistent with previous
theoretical results using the HSE functional, which yielded
1.37~eV~\cite{thiering2021niv}.

\subsection{Coupling parameters}

\begin{figure}
  \centering
  \includegraphics[width=\linewidth]{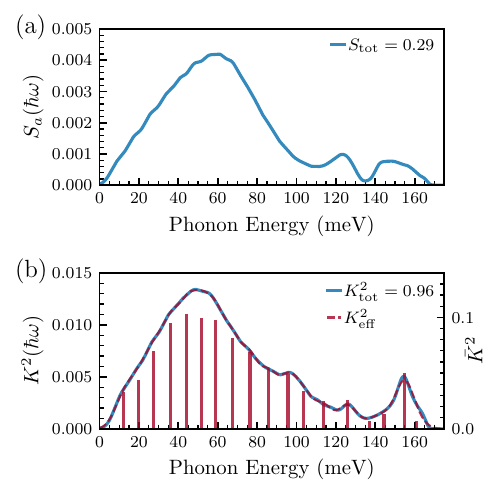}
  \caption{Spectral densities for (a) HR electron--phonon
    coupling $S(\hbar\omega)$ in units of $1/\text{meV}$, associated with
    $a_{1g}$-symmetry phonons, and (b) JT linear coupling
    $K^2(\hbar\omega)$. In panel (b), red stems represent the effective
    modes with their respective frequencies $\omega_n$ and coupling
    strengths $\bar{K}^2_n$. The red dashed line illustrates the
    spectral density obtained using the effective mode approximation
    $K^2_{\mathrm{eff}}(\hbar\omega)=\sum_{n=1}^{N_{\mathrm{eff}}}
    \bar{K}^2_{n}g_{\sigma}(\hbar\bar{\omega}_{n}-\hbar\omega)$. Both
    panels also include the total parameters $S_{\text{tot}}$ and
    $K^2_{\text{tot}}$.\label{fig:coupling}}
\end{figure}

To obtain the vibrational structure of the ground state and relaxation profiles
required to compute electron--phonon coupling parameters for both HR and JT
couplings, we employ the embedding methodology described in Sec.~\ref{sec:vibdil}.
Details and parameters of this methodology, along with convergence tests, are provided in Sec.~1 of the
Supplemental Material~\cite{supp}. This approach enabled the modeling of a large
$18\times 18\times 18$ supercell, which comprises 46\,655 atoms. We calculated
the spectral densities for two distinct types of interactions: the coupling for
$a$-symmetry modes, represented by $S(\hbar\omega)$, and the JT coupling for
$e$-symmetry modes, denoted by $K^2(\hbar\omega)$. To achieve a smooth
description of the electron--phonon interaction, the delta functions were
approximated using Gaussian functions with a variable width, $\sigma$, which
linearly decreases from 3.6~meV at $\omega = 0$ to 1.5~meV at the highest energy
phonon. Our results are illustrated in Fig.~\ref{fig:coupling}. Panel (a) shows
the spectral density $S(\hbar\omega)$ for the HR electron--phonon coupling in
units of $1/\text{meV}$, specifically linked to $a_{1g}$-symmetry vibrational
modes. Panel (b) depicts the spectral density for the JT linear coupling
$K^2(\hbar\omega)$. Figures~\ref{fig:coupling}(a) and (b) also include values for cumulative metrics: the
total HR parameter $S_{\text{tot}} = \sum_k S_k$ and the total JT coupling
$K^2_{\text{tot}} = \sum_k K^2_k$.

Both $S_k$ and $K^2_k$ define changes in the adiabatic potential
energy surface and are directly linked to relaxation energies. The
expression $\Delta E_{a} = \sum_k \hbar \omega_k S_k$ represents the
relaxation energy along the symmetry-preserving direction consequent to
the vertical transition \mbox{$A \to E$}. Conversely,
$\Delta E_{\text{JT}} = \sum_k \hbar\omega_k K^2_k$ accounts for the
JT relaxation, which describes the system's progression
along the $e$-symmetry direction from a high-symmetry to a
low-symmetry lowest-energy configuration. Our calculations yield
$\Delta E_{a} = 18.8$~meV and $\Delta E_{\text{JT}} = 63.5$~meV,
indicative of a highly pronounced JT contribution to the overall
electron--phonon interaction.

\subsection{Spectral functions}

\begin{figure}
  \centering
  \includegraphics[width=1\linewidth]{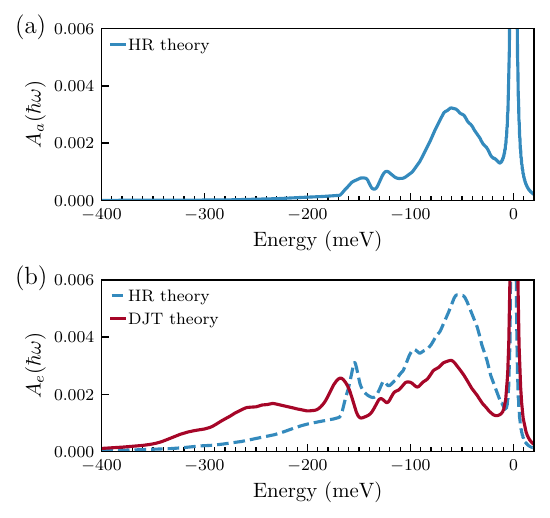}
  \caption{(a) The spectral function $A_a(\hbar\omega)$ for
    $a$-symmetry modes; (b) the spectral function $A_{e}(\hbar\omega)$
    for $e$-symmetry modes. In panel (b), we compare spectral
    functions derived from the dynamical Jahn--Teller theory (solid red
    line) and the Huang--Rhys theory (dashed blue line).\label{fig:A}}
\end{figure}

\begin{figure}
  \centering
  \includegraphics[width=8.5cm]{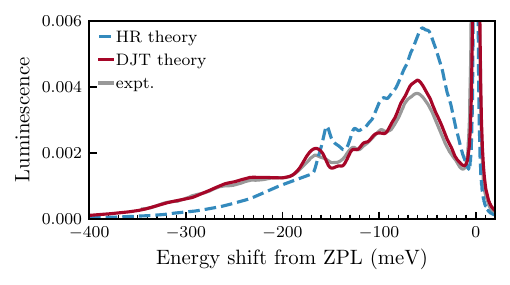}
  \caption{Theoretical normalized luminescence lineshapes (in units of
  1/meV) calculated using Huang--Rhys theory (dashed blue line) and
  dynamical Jahn--Teller theory (solid red line), juxtaposed
  with the experimental spectrum from Ref.~\onlinecite{collins1983}, recorded at
  77~K.\label{fig:theory_lum}}
\end{figure}

The spectral function $A_{a}(\hbar\omega)$ for $a$-symmetry modes, shown in
Fig.~\ref{fig:A}(a), was calculated using the generating function approach
[Eq.~\eqref{eq:generating-lum}]. The $\gamma$ parameter was set to
$0.35$~meV to match the linewidth of the experimental zero-phonon line (ZPL).
Notably, $A_{a}(\hbar\omega)$ exhibits a rapid decay to lower energies, becoming
negligible for spectral features more than 170~meV below the ZPL. This result
contrasts with the experimental lineshape extending to 300~meV below the ZPL
\cite{collins1983} (see Fig.~\ref{fig:theory_lum}), indicating that
symmetry-preserving modes alone cannot explain these experimental features.

We calculated the spectral function $A_e(\hbar\omega)$ for $e$-symmetry modes
using two distinct methodologies, as depicted in Fig.~\ref{fig:A}(b). In the HR
approach, $e$ modes were treated analogously to symmetric $a$-modes. This
involves interpreting $\Delta \mathbf{Q}_{\mathrm{JT}}$ as a displacement
between two harmonic potentials, a method previously employed for
NiV$^-$~\cite{thiering2021niv}. In contrast, the DJT treatment begins with the
diagonalization of the vibronic Hamiltonian
$\hat{\mathcal{H}} = \hat{\mathcal{H}}_{0} + \hat{\mathcal{H}}_{\mathrm{JT}}$
[refer to Eqs.~\eqref{eq:matrix_element2} and~\eqref{eq:matrix_element}] using a
selected set of effective modes. For the NiV$^-$ case, it was determined that a
count of $N_{\mathrm{eff}}=18$ effective modes, indicated by red vertical lines
in Fig.~\ref{fig:coupling}(b), suffices to achieve a spectral resolution of
$\sigma=5$~meV. Further convergence details are provided in Sec.~2 of the
Supplemental Material~\cite{supp}. Following the calculation of the vibronic
states [Eq.~\eqref{eq:1}], we estimate the overlap integrals contributing to
$A_{e}(\hbar\omega)$ [Eq.~\eqref{eq:spectral_f}] and approximate delta
functions with Gaussian functions having a width of $5$~meV.

Figure~\ref{fig:A}(b) clearly shows that the JT treatment yields substantially different results compared to the
HR function. The DJT spectral function extends more than 300~meV below the ZPL,
whereas HR theory produces pronounced features up to 170~meV below the ZPL but
decays below this energy. It~is evident that DJT theory is crucial for
accurately describing the overall extent of the experimental lineshape.

\subsection{Luminescence lineshapes}

The final luminescence lineshapes, derived from $A_a(\hbar\omega)$ and
$A_e(\hbar\omega)$ using Eqs.~\eqref{eq:P}, are showcased in
Fig.~\ref{fig:theory_lum} alongside the experimental curve from
Ref.~\cite{collins1983}, recorded at 77~K. Although the lineshape derived from
HR theory is consistent with the findings of Ref.~\cite{thiering2021niv}, it
fails to capture the experimental trend, underscoring the limitations of the HR
approach in describing optical features associated with dominant JT interactions.
In contrast, the lineshape obtained from the multimode DJT method (solid red
curve in Fig.~\ref{fig:theory_lum}) shows excellent agreement with the
experimental spectrum, accurately capturing all spectral features as well as the
intensity redistribution. By integrating the ZPL region (above $-15$~meV), the
Debye--Waller factor is estimated to be 35\% for the HR theory and 41\% for the
DJT theory. The DJT approach thus predicts a higher ZPL intensity, with a
sideband that is more extended compared to that predicted by HR theory.

As an extension to our investigation of phonon sidebands, we employed the
multi-mode vibronic solution detailed in Eq.~\eqref{eq:1} to calculate the Ham
factor $p$~\cite{ham1965} for the lowest vibronic state, which influences
spin--orbit interaction in DJT systems. This factor quantifies the reduction in
purely electronic spin--orbit splitting, $\lambda_0$. Under the assumption of
stationary ions, the intrinsic splitting $\lambda_0$ is estimated for electronic
orbitals. In contrast, the experimentally observed coupling derived from
vibronic states is represented by $\lambda = p\lambda_0$. The formula for $p$ is
given by
$p = \sum_{\mathbf{nl}} |C^{+}_{f;m\mathbf{nl}}|^2 - \sum_{\mathbf{nl}} |C^{-}_{f;m\mathbf{nl}}|^2$~\cite{thiering2018}.
We compute a value $p = 0.157$. The result obtained with the single-mode
approximation~\cite{thiering2021niv} was $p = 0.124$. The similarity in these
values indicates that, at least for the case of the NiV$^-$ center in diamond, a
multi-mode analysis may not be crucial for effectively capturing the spin-orbit
quenching.

\section{Conclusions}

In this study, we addressed the challenge of accurately modeling systems where
Jahn--Teller (JT) interactions play a dominant role. Utilizing a multi-mode
dynamical JT (DJT) approach, we were able to capture the complex
optical lineshapes observed in systems characterized by strong JT interactions,
specifically in the negatively charged nickel-vacancy center (NiV$^{-}$) in
diamond.

The methodology developed in Ref.~\onlinecite{razinkovas2021}, which was
initially applied to a case where JT effects were relatively weak, has now been
rigorously tested and validated in a scenario where these interactions have a
pronounced impact on spectral features. The findings confirm that a multi-mode
approach is crucial for reproducing accurate optical lineshapes and matching
experimental observations.

Looking ahead, the implications of this refined JT methodology extend beyond
merely reproducing spectral lineshapes. Future research should apply the
methodology to {\em predicting} spectra, and explore how the approach can
describe other electron--phonon related processes, such as nonradiative rates
and temperature broadening of the zero-phonon line (ZPL). These directions
promise to further our understanding of JT effects, potentially leading to
enhanced control and manipulation of electronic and optical properties of
defects and impurities in semiconductors as well as emitters and qubits applied
in quantum technologies.

\section*{Acknowledgments}

R.S., V.\v{Z}., G.T., A.G., A.A., and L.R. were supported by the QuantERA grant
{\sc SensExtreme}, funded by the Lithuanian Research Council (Grant No.
S-QUANTERA-22-1) and the National Office of Research, Development and Innovation of Hungary
(NKFIH Grant No. 2019-2.1.7-ERA-NET-2022-00040).
A.G. acknowledges the support from Quantum Information National Laboratory of Hungary
(NKFIH Grant No. 2022-2.1.1-NL-2022-00004) and European Commission for the projects QuMicro
(Grant No. 101046911) and SPINUS (Grant No. 101135699).
C.~VdW. was supported by the U.S. Department of
Energy, Office of Science, National Quantum Information Science Research
Centers, Co-design Center for Quantum Advantage (C2QA) under contract number
DE-SC0012704. Computations were performed on the supercomputer {\sc GALAX} of
the Center for Physical Sciences and Technology, Lithuania, and on the High
Performance Computing Center ``HPC Saul{\.e}tekis'' in the Faculty of Physics,
Vilnius University.


\newpage
\input{main.bbl}


\end{document}


\title{Supplemental Material for:\\
  Optical lineshapes for orbital singlet to doublet transitions in a dynamical Jahn--Teller system: the NiV$^{-}$ center in diamond}

\author{Rokas Silkinis}
\author{Vytautas \v{Z}alandauskas}
\affiliation{Center for Physical Sciences and Technology (FTMC),
  Vilnius LT-10257, Lithuania}

\author{Gerg\H{o} Thiering}
\affiliation{Institute for Solid State Physics and Optics,
  HUN-REN Wigner Research Centre for Physics,
  P.O.~Box~49, H-1525 Budapest, Hungary}

\author{Adam Gali}
\affiliation{Institute for Solid State Physics and Optics,
  HUN-REN Wigner Research Centre for Physics,
  P.O.~Box~49, H-1525 Budapest, Hungary}
\affiliation{Department of Atomic Physics, Institute of Physics,
  Budapest University of Technology and Economics,
  M\H{u}egyetem rkp.~3,
  H-1111 Budapest, Hungary}
\affiliation{MTA-WFK Lend{\"u}let ``Momentum''
  Semiconductor Nanostructures Research Group,
  PO.~Box~49, H-1525 Budapest, Hungary}

\author{Chris G. Van de Walle}
\affiliation{Materials Department, University of California,
  Santa Barbara, CA~93106-5050, USA}

\author{Audrius Alkauskas}
\thanks{Deceased}
\author{Lukas Razinkovas}
\email{lukas.razinkovas@ftmc.lt}
\affiliation{Center for Physical Sciences and Technology (FTMC),
  Vilnius LT-10257, Lithuania}

\maketitle

\vspace{-1cm}
\section{Details of the embedding methodology}

The vibrational structure of large supercells was analyzed using the force
constant embedding methodology, described in
Refs.~\cite{alkauskas2014,razinkovas2021}. This method utilizes the short-range
nature of interatomic interactions in semiconductors to construct a Hessian
matrix for supercells containing thousands of atoms. The criteria for
constructing Hessian matrix elements are as follows: for pairs of atoms within a
cutoff radius $r_{d}$ from any vacancy or the Ni atom, matrix elements from the
actual $4\times4\times4$ defect supercell are used. Otherwise, if two atoms are
within another specified cutoff radius $r_{b}$, elements from the Hessian matrix
of the bulk $4\times4\times4$ supercell are used. Matrix elements are set to
zero in all other cases. The cutoff radii selected for this embedding procedure
are $r_{d} = 5.58$~{\AA} and $r_{b} = 6$~{\AA}. The computational parameters
employed are consistent with those presented in the main text.

\begin{figure}
  \centering
  \includegraphics[width=0.55\linewidth]{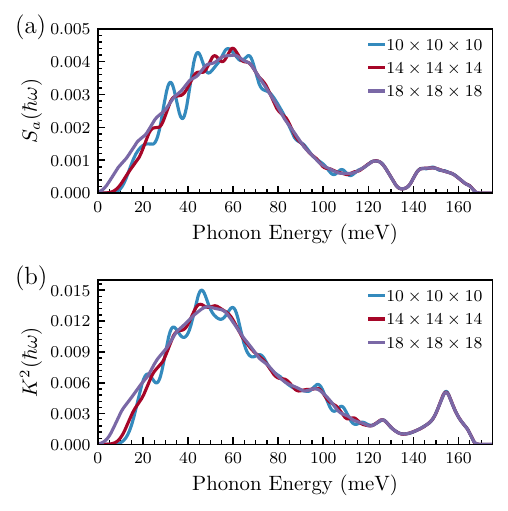}
  \caption{Convergence of the spectral densities of electron--phonon coupling
    with respect to the supercell size
    for (a) $S_{a}(\hbar\omega)$ associated with $a$-symmetry phonons, and (b)
    $K^{2} (\hbar\omega)$ associated with $e$-symmetry phonons.
    Both spectral densities are expressed in units of meV$^{-1}$.
    Supercells range in size from $10\times10\times10$ (8000 atomic sites) to
    $18\times18\times18$ (46656 atomic sites).}
  \label{fig:S_tot_conv}
\end{figure}

We evaluated the convergence of electron--phonon spectral densities for $a$- and
$e$-symmetry modes. For symmetry-preserving $a$ modes, we calculated the
Huang--Rhys (HR) spectral density
$S_{a}(\hbar\omega) = \sum_k S_k\delta(\hbar\omega_{k} - \hbar\omega)$, depicted
in Fig.~\ref{fig:S_tot_conv}(a), where $S_{k}$ represents the partial HR
parameters~\cite{Huang1950}. Figure~\ref{fig:S_tot_conv}(b) illustrates the
convergence of spectral density for Jahn--Teller (JT) active modes, expressed as
$K^2(\hbar\omega) = \sum_{k}K^{2}_{k}\delta(\hbar\omega_{k} - \hbar\omega)$,
with $K_{k}$ denoting dimensionless vibronic parameters~\cite{obrien1972}.
Supercell sizes of $10 \times 10 \times 10$ (8000 atomic sites),
$14 \times 14 \times 14$ (21952 atomic sites), and $18 \times 18 \times 18$
(46656 atomic sites) were tested. To achieve a smooth representation of the
electron--phonon spectral function, we approximated delta functions with Gaussian
functions of variable widths ($\sigma$), decreasing linearly from 3.6~meV at zero
frequency to 1.5~meV at the highest phonon energy. Convergence tests confirm that
the $18 \times 18 \times 18$ supercell size provides a comprehensive and smooth
spectral density of electron--phonon coupling for selected smoothing parameters,
effectively capturing contributions from both low-energy acoustic and
higher-frequency phonon modes.

\section{Benchmarking the diagonalization of the Jahn--Teller Hamiltonian}

\begin{figure}
  \centering
  \includegraphics[width=0.55\linewidth]{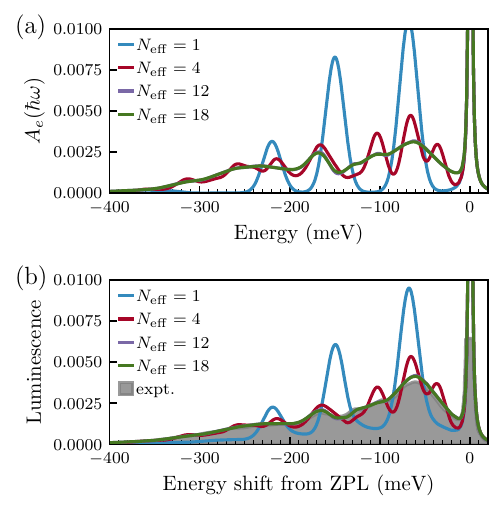}
  \caption{
    (a) Convergence of the theoretical spectral function $A_{e}(\hbar\omega)$ (in units of meV$^{-1}$) calculated using dynamical Jahn--Teller theory with respect to the number of effective modes $N_{\mathrm{eff}}$ for $e$-symmetry modes.
    (b) Convergence of the theoretical normalized luminescence lineshapes (in units of meV$^{-1}$) calculated using dynamical Jahn--Teller theory with respect to the number of effective modes $N_{\mathrm{eff}}$.
    The total number of excited phonons used for each lineshape was $n_{\mathrm{tot}}=5$.
    Experimental spectrum from Ref.~\cite{collins1983}, recorded at 77 K.
    \label{fig:N_eff}
  }
\end{figure}

We analyzed the convergence of computed lineshapes concerning the parameters
that define the basis of the JT Hamiltonian, described by
Eqs.~(11) and (12) of the main text. Specifically, we investigated how both the
optical spectral function $A_{e}(\hbar\omega)$, defined in Section III of the
main text, and the calculated lineshape $L(\hbar\omega)$, outlined in Eq.~(4) of
the main text, vary with different numbers of effective modes,
$N_{\mathrm{eff}}$.
Figure~\ref{fig:N_eff}(a) depicts $A_{e}(\hbar\omega)$, which pertains exclusively
to JT-active modes, calculated for $N_{\mathrm{eff}} = 1, 4, 12$, and $18$.
The effective parameters for these calculations are listed in Table~\ref{tab:K2}.

\newpage

The parameters for a single effective mode ($N_{\mathrm{eff}} = 1$) were
calculated according to the formulations provided by M.C.M
O'Brien~\cite{obrien1972}:
\begin{equation}
  \hbar\omega_{\text{eff}} = \frac{\sum_{k}\hbar\omega_{k}K_{k}^{2}}{\sum_{k}K_{k}},
  \qquad k_{\text{eff}} = \sum_{k}K_{k}^{2},
\end{equation}
where $k$ encompasses all $e$-symmetry modes within the $18 \times 18 \times 18$
supercell. Figure~\ref{fig:N_eff}(b) displays the actual lineshape, obtained by
convolving $A_{e} (\hbar\omega)$ [calculated for different $N_{\mathrm{eff}}$,
as shown in Fig.~\ref{fig:N_eff}(a)] with the spectral function
$A_{a} (\hbar\omega)$ for $a$-symmetry modes, depicted in Fig.~3(a) of the main
text. We compared these calculated lineshapes with experimental
spectrum~\cite{collins1983}, finding that a single effective mode does not
adequately capture the spectral features of this multi-mode JT system. Our
results indicate that $N_{\mathrm{eff}} = 12$ effectively reproduces the
experimental spectrum; however, to enhance accuracy, we employed a higher number
of effective modes, $N_{\mathrm{eff}} = 18$.

\begin{figure}
  \centering
  \includegraphics[width=0.55\linewidth]{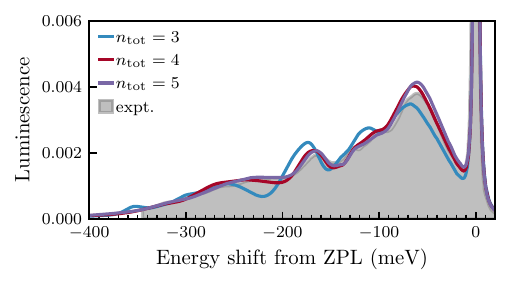}
  \caption{
    Convergence of the theoretical normalized luminescence lineshapes (in units of meV$^{-1}$) calculated using dynamical Jahn--Teller theory with respect to the total number of excited phonons $n_{\mathrm{tot}}$.
    The number of effective modes used for each lineshape was $N_{\mathrm{eff}}=18$.
    Experimental spectrum from Ref.~\cite{collins1983}, recorded at 77 K.
    \label{fig:n_max}
  }
\end{figure}

Figure~\ref{fig:n_max} illustrates the convergence of the theoretical normalized
luminescence lineshapes with respect to the total number of excited phonons
$n_{\mathrm{tot}}$ included in constructing the phonon basis, revealing that
using $n_{\mathrm{tot}} = 5$ excited phonons leads to a converged result for the
NiV$^{-}$ center in diamond.

\begin{table}
\centering
\caption{Effective parameters of a multi-mode Jahn--Teller system for different $N_{\mathrm{eff}}$}
\label{tab:K2}
\begin{tabular}{ccccccccc}
\toprule
& \multicolumn{2}{c}{$N_{\mathrm{eff}} = 1$} & \multicolumn{2}{c}{$N_{\mathrm{eff}} = 4$} & \multicolumn{2}{c}{$N_{\mathrm{eff}} = 12$} & \multicolumn{2}{c}{$N_{\mathrm{eff}} = 18$} \\
\cmidrule(lr){2-3} \cmidrule(lr){4-5} \cmidrule(lr){6-7} \cmidrule(lr){8-9}
Nr.  & Frequency (meV) & $k^2$ & Frequency (meV) & $k^2$ & Frequency (meV) & $k^2$ & Frequency (meV) & $k^2$ \\
\midrule
1  &66.34 & 0.956& 58.504  & 0.367 & 15.034  & 0.058 & 12.146  & 0.033 \\
2  &&& 32.869  & 0.300 & 26.960  & 0.094 & 19.913  & 0.044 \\
3  &&& 93.642  & 0.178 & 37.843  & 0.126 & 27.570  & 0.070 \\
4  &&& 147.386 & 0.100 & 48.604  & 0.150 & 35.909  & 0.095 \\
5  &&&         &       & 59.626  & 0.142 & 44.173  & 0.103 \\
6  &&&         &       & 71.834  & 0.102 & 51.664  & 0.099 \\
7  &&&         &       & 84.106  & 0.072 & 59.096  & 0.098 \\
8  &&&         &       & 97.402  & 0.071 & 67.411  & 0.081 \\
9  &&&         &       & 110.328 & 0.034 & 76.371  & 0.069 \\
10 &&&         &       & 125.245 & 0.030 & 85.963  & 0.054 \\
11 &&&         &       & 141.048 & 0.013 & 95.701  & 0.049 \\
12 &&&         &       & 154.796 & 0.063 & 103.646 & 0.034 \\
13 &&&         &       &         &       & 113.648 & 0.025 \\
14 &&&         &       &         &       & 125.927 & 0.026 \\
15 &&&         &       &         &       & 136.922 & 0.007 \\
16 &&&         &       &         &       & 144.417 & 0.013 \\
17 &&&         &       &         &       & 154.609 & 0.050 \\
18 &&&         &       &         &       & 161.080 & 0.007 \\
\bottomrule
\end{tabular}
\end{table}

\input{supplemental.bbl}

%% file: main.bbl
%